# Forecasting ability of a multi-renewal seismicity model for Italy

G. Molchan and L. Romashkova

**Summary.** The inter-event time (IET) is sometimes used as a basis for prediction of large earthquakes. It is the case when theoretical analysis of prediction is possible. Quite recently a specific IET- model was suggested for dynamic probabilistic prediction of $M \geq 5.5$ events in Italy (http://earthquake.bo.ingv.it). In this study we analyze both some aspects of the statistical estimation of the model and its predictive ability. We find that more or less effective prediction is possible within 4 out of 34 seismotectonic zones where seismicity rate or clustering of events is relatively high. We show that, in the framework of the model, one can suggest a simple zone independent strategy, which practically optimizes the relative number of non-accidental successes, or the Hanssen-Kuiper ($HK$) skill score. This quasi-optimal strategy declares alarm in a zone for the first 2.67 years just after the occurrence of each large event in the zone. The optimal $HK$ skill score values are: ≈26% for the 3 most active zones and 2-10% for the 26 least active zones. However, the number of false alarm time intervals per one event in each of the zones is unusually high: ≈0.7 and 0.8-0.95 respectively. Both these theoretical estimations are important because any prospective testing of the model is unrealistic in most of the zones during a reasonable time. This particular analysis requires a discussion of the following issues of general interest: a specific approach to the analysis of predictions vs. the standard CSEP testing approach; prediction vs. forecasting; $HK$ skill score vs. probability gain; the total forecast error diagram and connected false alarms.

## 1. Introduction

A sequence of large earthquakes in a region is often considered as a renewal process, viz., inter-event times are assumed to be independent random variables with a distribution function $F(t)$, $t > 0$. If the distribution is not exponential (the case of a non-poissonian process), the model may possess nontrivial predictive ability. The issue was discussed by Molchan (1990, 1991, 1997, 2003) in a framework of the general theory of prediction optimization. There are some points to remind. The structure of optimal prediction for a renewal process is very simple when we measure the quality of prediction by an increasing loss-function, $\varphi(n, \tau)$ of two long-term parameters: the relative number of failures-to-predict $n$ and the relative alarm time $\tau$. The optimum of the measure $\varphi = \max(n, \tau)$ weakly depends on the type of distribution $F$ (Weibull, Gamma, or Log-Normal), when the coefficient of variation $I$, i.e., the ratio of the standard error $\sigma_F$ to the mean $m_F$, is fixed. For example, assuming $I < 0.6$ (the realistic case for characteristic earthquakes on the San Andreas fault), one has $\varphi \leq 0.35$. The meaning of this relation is the following: the total alarm duration of 35% yields 35% of accidental successes plus additional 30% of non-accidental ones (i.e., $1 - n - \tau$). For comparison, the essentially more complicated M8-algorithm has respectively $\approx 32 + 23\%$ successes in the real-time prediction of $M = 8.0 \div 8.5$ earthquakes worldwide (Molchan & Romashkova, 2010).

To predict characteristic earthquakes in the San Andreas fault, the Working Group on California Earthquake Probabilities (1988) used the renewal model with $F$ that is specific to the fault segment concerned. In this case one deals with the *multi-renewal model* such that all $F$ belong to the Log-Normal family of distributions and differ by the parameters $m_F$, $\sigma_F$. The choice of the distribution type for large earthquakes is a serious problem because of scarcity of our knowledge relative to the behavior of $F$ at large $t$. To overcome the difficulty, Faenza et al. (2003) draw attention to the *Cox proportional-hazards regression model* (Kalbfleisch & Prentice, 2003) where the problem of the $F$-model including the tail behavior is solved empirically. The Cox model adapted to the forecasting of $M \geq 5.5$ events in Italy supposes the renewal property in each of 34 zones with a zone-specific distribution $F_i$. In order to achieve efficiency in the estimation of $\{F_i\}$ the renewal processes in different zones are assumed to be independent, and the survival functions, $1 - F_i(t)$, belong to the one-parametric (Lehmann) family:

$$1 - F_i(t) = (1 - F_0(t))^{\kappa_i}. \qquad (1)$$

The unknown basic distribution $F_0$ and exponents $\kappa_i$ are fitted using the real data.

This model is specified in Cinti et al. (2004) and is used for dynamic probabilistic prediction in Italy (http://earthquake.bo.ingv.it).

The goal of our study is:

- to investigate the predictive ability of the multi-renewal model assuming it to be true;
- to discuss some statistical aspects of the model (1); in particular, the role of regionalization in the model, since the zone portioning is not unique.

The analysis of the first point is based on Molchan & Keilis-Borok (2008) and on Molchan (2010) where the space-time prediction problem is discussed in detail.

## 2. Background

The concepts and notions of this study are described in detail in Molchan (1997, 2010). We assume that the sequence of target events in an area $G$ follows a renewal model with distribution F(s) = Pr(inter-event time<s). The instantaneous conditional rate of an event at a moment $t$ is then given by the *hazard function*

$$h(s) = F'(s)/(1 - F(s)) = -(d/ds)\ln(1 - F(s)), \qquad (2)$$

provided that $t - s$ is the time of the last event in $G$. For this reason the prediction strategy that minimizes an increasing convex loss function $\varphi(n, \tau)$ looks as follows: the alarm is realized at the moment $t$ if $h(s) > p_\varphi$, where the threshold $p_\varphi$ is a functional of $\varphi$.

The strategy as described above involves two errors:

- the relative rate of failures-to-predict

$$n(p) = \int_0^\infty [h(s) < p]dF(s), \qquad (3)$$

- the rate of alarm time

$$\tau(p) = \lambda \int_0^\infty [h(s) > p](1 - F(s))ds, \qquad (4)$$

where [A] is the logical 1-0 function; $\lambda$ is a stationary rate of target events in $G$ and therefore

$$\lambda^{-1} = \int_0^\infty s dF(s) = \int_0^\infty (1 - F(s)) ds \,. \tag{5}$$

The curve $\Gamma = \{(n(p), \tau(p)), p > 0\}$ gives the *error diagram; n* as a function of $\tau$ is decreasing and convex. The diagram characterizes the predictive ability of the renewal model, since for any nontrivial prediction strategy, the error couple $(n, \tau)$ as a point belongs to the convex hull of two curves: $\Gamma$ and diagonal $n + \tau = 1$. The diagonal represents the errors of random guess strategies, while any point of $\Gamma$ represents the errors of the optimal strategy that minimizes some loss function $\varphi(n, \tau)$. For example, the point of $\Gamma$ that is the farthest from the diagonal maximizes the quantity

$$HK = 1 - n - \tau, \quad (\textit{Hanssen-Kuiper skill score}), \tag{6}$$

i.e., minimizes $\varphi(n, \tau) = n + \tau$.

The quantity (6) gives the expected relative rate of events whose prediction is not accidental. For this reason $HK$ is a measure of efficiency for the prediction strategy in hand. The maximum value of (6) is attained when the alarm rule with threshold $p_\varphi = \lambda$ is used, viz.,

$$\text{alarm: } h(s) > \lambda \quad (\textit{HK-optimal strategy}) \,. \tag{7}$$

In the more general case: $\varphi = n + a\tau$, the optimal strategy is given by the following threshold:

$$p_\varphi = a\lambda, \quad \varphi(n, \tau) = n + a\tau \,. \tag{8}$$

The above statements are applicable to a multi-renewal model. In this model
- the space is divided into zones, $G = \bigcup G_i$;
- the target events in different zones are independent;
- the process in each zone follows a renewal model with distribution $F_i$ and rate $\lambda_i$.

The function (6) is still an exact measure of efficiency if $(n, \tau) = (n_\Sigma, \tau_\Sigma)$, where

$$n_\Sigma = \sum \lambda_i n_i / \sum \lambda_i, \quad \tau_\Sigma = \sum \lambda_i \tau_i / \sum \lambda_i, \tag{9}$$

and $(n_i, \tau_i)$ are the errors of prediction in $G_i$. The efficiency becomes the highest if the alarm rule (7) is applied in each zone with specific characteristics $h(s)$ and $\lambda$.

Usually a forecasting is represented by probabilities to have one or more target events in zones $\{G_i\}$ for the period $(t, t+\Delta)$, $\{p_i(t|\Delta)\}$. Thereby we introduce a common *probabilistic scale of hazard* for these zones.

The situation in prediction (0-1 expression) is different because here we use the so-called *probability gain scale*. In the case of the multi-renewal model this can be illustrated as follows. One has

$$p_i(t|\Delta) = [F_i(s+\Delta) - F(s)]/[1 - F_i(s)], \qquad (10)$$

where $t-s$ is the time of the last target event in zone $G_i$. By (2), $p_i(t|\Delta) \approx h_i(s)\Delta$, $\Delta \ll 1$. To optimize $\varphi(n,\tau) = n + a\tau$, we consider a sequence of small intervals $\Delta$ and compare $p_i(t|\Delta)$ with $a\lambda_i\Delta$ (see (8)). But $\lambda_i\Delta = p_i^0(\Delta)$ is the stationary probability of target event in $\Delta$. Therefore the hazard of the zone is characterized by the normalized probability $p_i/p_i^0$ that is the local probability gain.

To define the forecasting ability of a multi-renewal model, we consider a standardized alarm rule, viz.,

$$\text{alarm in } G_i: \quad h_i(s) > p \qquad (\textit{forecasting strategy}) \qquad (11)$$

in contrast to the *HK* optimal rule (7) where $p = \lambda_i$.

Strategy (11) is characterized by the *total error diagram* $\Gamma_f = \{(n_\Sigma(p), \tau_\Sigma(p)), p > 0\}$, where

$$n_\Sigma(p) = \sum \lambda_i n_i(p) / \sum \lambda_i, \quad \tau_\Sigma(p) = \sum \lambda_i \tau_i(p) / \sum \lambda_i, \qquad (12)$$

and $n_i(p), \tau_i(p)$ are determined by (3,4) with $F = F_i$. This diagram is not convex in contrast to one's zonal analogue.

*Connected alarms.* Suppose that the alarm time set within an inter-event time is a collection of isolated intervals. Any interval that does not culminate in target event is considered as a "*false (connected) alarm*". The number of the false alarm intervals per one event, $n_A$, is of interest for applications. It is easy to derive $n_A$ for any renewal model. The simplest solution that will be interesting later on is as follows:

If $h_i(s)$ is unbounded at 0 and has *U*-shape (more precisely, any set $\{s: h_i(s) < p\}$ is an interval $(a_i^-(p), a_i^+(p))$), then $n_{A,i}(p) = 1 - F_i(a_i^-(p))$ for strategy (11).

By (3), $n_{A,i}(p) \geq n_i(p)$ and

$$n_{A,i}(p) = n_i(p) \qquad \text{if} \qquad a_i^+(p) = \infty. \qquad (13)$$

For the multi-renewal model one has

$$n_{A\Sigma}(p) = \sum \lambda_i n_{A,i}(p) / \sum \lambda_i \qquad (14)$$

provided that the connected alarms in different zones are calculated separately.

**3. The multi-renewal model for Italy.**

*The model.* According to Cinti et al. (2004), the seismoactive area of Italy is divided into 34 tectonically uniform zones (Fig 1); the distributions $F_i$ in these zones are estimated from the assumption of proportionality of the hazard functions: $h_i(s) = \kappa_i h_0(s)$ (see (1,2)), using the inter-event time data represented in Table 1.

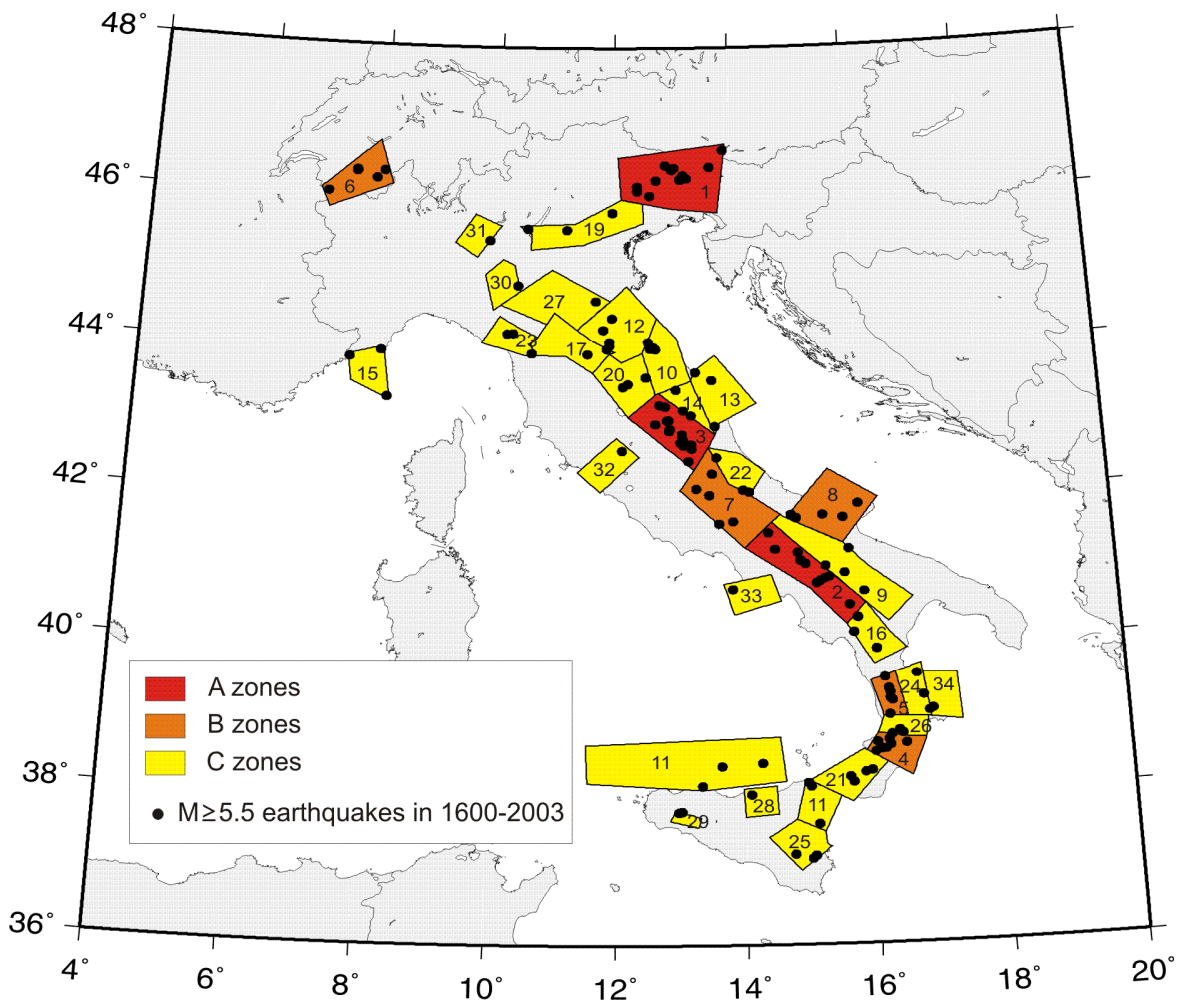

**Figure 1**. Seismotectonic zonation of Italy according to Cinti et al (2004). Only zones that have experienced at least one $M \geq 5.5$ event in 1600-2003 are shown. Fore description of A, B and C zones see the text.

Figure 3 from Cinti et al. (2004) contains information on the basic survival function

$$1 - F_0(s) = \exp\{-\int_0^s h_0(u)du\} \tag{15}$$

in the form of an empirical curve $\varepsilon : \{\ln(s^{-1}\int_0^s h_0(u)du) \text{ vs. } \lg s\}$ The plot of $\varepsilon\{\cdot\}$ is linear with a negative slope up to $s = t_+$ and degenerates to the 0-line, $\varepsilon : \{0 \text{ vs. } \lg s\}$ for $s \geq t_+$. Hence, we have approximately

$$\int_0^s h_0(u)du = t_+^{\gamma} s^{1-\gamma}[s \leq t_+] + s[s > t_+], \tag{16}$$

where $\gamma \approx 0.8$ and $t_+ \approx 10^{1.3} \approx 20 \, years$.

**Table 1.** Inter event time (IET) data in years for $M \geq 5.5$ earthquakes in Italy for the period $T$=1600-2003 according to Cinti et al (2004).

| Zones of group A | | | Zones of group B | | | | |
|---|---|---|---|---|---|---|---|
| 1 | 2 | 3 | 4 | 5 | 6 | 7 | 8 |
| 90.3+ | 88.4+ | 39.8+ | 59.8+ | 38.2+ | 99.6+ | 54.6+ | 27.6+ |
| 10.3 | 6.3 | 0.02 | 84.1 | 129.3 | 49.8 | 108.2 | 0.0 |
| 76.0 | 7.5 | 63.2 | 39.2 | 68.2 | 40.7 | 141.4 | 0.02 |
| 12.3 | 30.7 | 0.01 | 8.7 | 18.3 | 0.34 | 10.9 | 0.08 |
| 24.0 | 72.7 | 0.05 | 113.9 | 16.6 | 57.6 | 69.3 | 18.7 |
| 60.7 | 20.5 | 27.3 | 22.5 | 42.7 | 99.6+ | 19.7+ | 129.5 |
| 54.7 | 27.2 | 16.9 | 19.2 | 90.5+ | | | 226.9 |
| 8.6 | 57.2 | 4.3 | 56.6+ | | | | 0.00 |
| 39.5 | 52.2 | 80.5 | | | | | 1.2+ |
| 0.35 | 0.0 | 6.1 | | | | | |
| 0.01 | 18.3 | 21.5 | | | | | |
| 0.00 | 23.1+ | 120.1 | | | | | |
| 21.6 | | 18.0 | | | | | |
| 5.7+ | | 0.00 | | | | | |
| | | 0.05 | | | | | |
| | | 6.2+ | | | | | |
| Zones of group C | | | | | | | | |
| 9 | 10 | 11 | 12 | 13 | 14 | 15 | 16 | 17 |
| 131.2+ | 187.0+ | 126.7+ | 61.2+ | 91.0+ | 141.3+ | 255.0+ | 186.2+ | 108.1+ |
| 120.4 | 88.2 | 253.7 | 27.1 | 239.8 | 58.3 | 32.2 | 31.9 | 128.8 |
| 78.9 | 41.2 | 22.3 | 93.0 | 12.9 | 73.6 | 76.4 | 160.1 | 21.1 |
| 59.8 | 0.25 | 1.3+ | 222.7+ | 60.2+ | 130.8+ | 40.5+ | 25.7+ | 146.0+ |
| 13.7+ | 87.4+ | | | | | | | |
| 18 | 19 | 20 | 21 | 22 | 23 | 24 | 25 | 26 |
| 108.1+ | 95.2+ | 181.4+ | 183.1+ | 106.8+ | 237.3+ | 38.4+ | 93.0+ | 26.3+ |
| 128.8 | 96.3 | 8.3 | 0.00 | 226.9 | 77.5 | 193.7 | 0.01 | 156.9 |
| 21.1 | 110.4 | 127.6 | 111.8 | 16.9 | 5.9 | 4.1 | 125.1 | 0.07 |
| 146.0+ | 102.2+ | 86.7+ | 14.1 | 53.3+ | 83.3+ | 167.7+ | 185.8+ | 220.8+ |
| | | | 95.0+ | | | | | |
| 27 | 28 | 29 | 30 | 31 | 32 | 33 | 34 | |
| 196.8+ | 223.2+ | 368.0+ | 371.5+ | 202.4+ | 95.4+ | 283.6+ | 332.0+ | |
| 207.2+ | 180.8+ | 0.0 | 32.5+ | 201.6+ | 308.6+ | 120.4+ | 72.0+ | |
| | | 36.0+ | | | | | | |

*Notation:* IET+ values are the censoring times; they correspond to the first and last earthquakes in the zones.

By (16), the hazard functions $h_i(s) = \kappa_i h_0(s)$ are U-shaped; the sets $\{s : h_i(s) < p\}$ are unbounded for $p > \kappa_i$. Therefore the previous statement about connected alarms is applicable to the Italian multi-renewal model. In particular,

$$n_{A,i}(p) = n_i(p), \qquad p > \kappa_i. \tag{17}$$

***Interpretation of*** $(\gamma, \kappa_i)$. Let us compare (16) with the analogous characteristic for a Poisson cluster model considered in Molchan (2005). This model consists of main events (a Poisson process of rate $\lambda^*$) and independent clusters (not necessarily Poissonian). Cluster events follow main ones with a rate $\lambda_{cl}(t)$ that is unbounded at 0. In these conditions we have the following asymptotic expressions for the inter-event time distribution $F$: $1 - F(t) \propto \exp\{-\lambda^* t\}$ at large times, and $h(t) \propto \lambda_{cl}(t)$ at small times. These expressions are in agreement with (1, 16) if $\lambda^* = \kappa$ and $\lambda_{cl}(t) \sim ct^{-\gamma}$, $t \to 0$. Therefore we can interpret $\kappa_i$ as the rate of "main" events in $G_i$, and $\gamma$ as the exponent that is present in Omori law for the cluster events in $\{G_i\}$. Note that the Omori law for large events has been studied insufficiently.

***Parameters*** $\kappa_i$. Relation (5) can be used to relate the unknown parameter $\kappa_i$ with the rate of target events in zone $G_i$:

$$\lambda_i^{-1} = \int_0^\infty (1 - F_i(s))ds = \int_0^\infty \exp(-\kappa_i \int_0^s h_0(u)du)ds. \tag{18}$$

Using (16) and the notation $\tilde{x}_i = \lambda_i t_+$, one has the following equation in $x = \kappa_i t_+$:

$$x/\tilde{x}_i = x \int_0^1 \exp(-x\tau^{1-\gamma})d\tau + e^{-x} \tag{19}$$

or $\quad x/\tilde{x}_i = \bar{\gamma} x^{1-\bar{\gamma}} \Gamma(x | \bar{\gamma}) + e^{-x}$, where $\bar{\gamma} = (1-\gamma)^{-1}$ and $\Gamma(x | \alpha) = \int_0^x u^{\alpha-1} e^{-u} du$ is the incomplete Gamma function.

Equation (19) has a unique solution. Hence, model (1) implies the identical seismic regime of target events in the zones with equal rates $\lambda_i$.

The right hand part of (19) is a decreasing function of $\gamma$, $0 < \gamma < 1$, and is equal to 1 at $\gamma = 0$. Hence, $x/\tilde{x}_i \leq 1$ or $\kappa_i < \lambda_i$. This relation appears very natural when $\kappa_i$ is interpreted as the rate of "main" events in $G_i$.

By (19), $x_i/\tilde{x}_i \sim 1$, if $\tilde{x}_i = \lambda_i t_+$ is small; more precisely

$$x_i/\tilde{x}_i \approx 1 - \tilde{x}_i^2 \gamma/(4-2\gamma) + \tilde{x}_i^3 \gamma/(9-6\gamma). \tag{20}$$

Combining the inequality $\lambda_i > \kappa_i$ with (17, $p = \lambda_i$), we conclude that the relative number of failures-to-predict $n$ and the relative number of false connected alarms $n_A$ are equal for the *HK* optimal strategy in $G = \bigcup G_i$.

***Error diagrams for the multi-renewal model.*** By (3, 4), the error diagram for alarm rule (11) in zone $G_i$ is specified as follows:

$$\Gamma_i = \{\tilde{n}_i(x_i q), \tilde{\tau}_i(x_i q)\}, \qquad (21)$$

where $q = (1-\gamma)/(t_+ p) > 0$ ( $q$ is treated as a new parameter) ;

$$\tilde{n}_i(u) = \exp\{-x_i u^{(1-\gamma)/\gamma}\} - [u > 1-\gamma]\exp(-x_i), \qquad 0 \le u \le 1 \qquad (22)$$

$$\tilde{\tau}_i(u) = \tilde{x}_i \bar{\gamma} x_i^{-\bar{\gamma}} \Gamma(x_i u^{(1-\gamma)/\gamma} \mid \bar{\gamma}) + [u > 1-\gamma]\tilde{x}_i x_i^{-1} \exp(-x_i), \quad 0 \le u \le 1 \qquad (23)$$

and $\tilde{n}_i(u) = 1 - \tilde{\tau}_i(u) = 0$ when $u > 1$. Remind that $x_i = \kappa_i t_+$, $\tilde{x}_i = \lambda_i t_+$, and $\bar{\gamma} = (1-\gamma)^{-1}$.

Using (12, 22, and 23), we can find the total error diagram of forecasting strategy (11) in $G = \bigcup G_i$ :

$$\Gamma_f = \{\sum \lambda_i \tilde{n}_i(q x_i)/\sum \lambda_i, \sum \lambda_i \tilde{\tau}_i(q x_i)/\sum \lambda_i\}, \quad q > 0, \qquad (24)$$

and the prediction errors of the *HK* optimal strategy:

$$n_\Sigma = \sum \lambda_i \tilde{n}_i(u_i)/\sum \lambda_i, \qquad \tau_\Sigma = \sum \lambda_i \tilde{\tau}_i(u_i)/\sum \lambda_i, \qquad u_i = (1-\gamma)x_i/\tilde{x}_i. \qquad (25)$$

***Quasi-optimal strategy.*** The alarm rule based on the relation $h_i(s) > \kappa_i$ is independent on the zone $G_i$ because $h_i(s)/\kappa_i = h_0(s)$. By (8), this rule is optimal relative to $\varphi(n, \tau) = n + (\kappa_i/\lambda_i)\tau$. For the Italian data one has $\kappa_i/\lambda_i \approx 1$ (see Table 2). Therefore we may expect that the strategy (11) with $p = \kappa_i$ is almost *HK* optimal.

From the practical point of view, the set $A = \{s: h_i(s) \ge \kappa_i\} = \{s: h_0(s) \ge 1\}$ is unstable because $h_0(s) = 1$ for any $s \ge t_+$. Therefore along with $A$ we consider the following family of sets

$$A_c = (o, t_-) \bigcup (t_+, t_+ + c), \qquad c \ge 0, \qquad t_- = (1-\gamma)^{1/\gamma} t_+ \qquad (26)$$

such that $A_c \subset A_\infty = A$.

The prediction errors for the alarm rules based on $A_0$ and $A_\infty$ can be found by substituting $u_i^\mp := (1-\gamma) \mp 0$ for $u_i = (1-\gamma)x_i/\tilde{x}_i$ in (25) (-0 and +0 are infinitesimal small numbers related to $A_0$ and $A_\infty$ respectively). Functions (22, 23) are continuous from the left and $x/\tilde{x}_i \le 1$. Therefore $u_i \to u_i^-$ as $x_i/\tilde{x}_i \to 1$, i.e., the errors for $A_0$ have to be similar to the *HK* optimal one provided that $x_i/\tilde{x}_i = \kappa_i/\lambda_i \approx 1$.

**Table 2**. Characteristics of the Italian multi-renewal model.

| Group | Zone | $N(T)$ | $\kappa/\lambda$ | $\lambda t_-$ | $n$ | $\tau$ | HK | PG | $\sigma/m$ |
|---|---|---|---|---|---|---|---|---|---|
| A | 1 | 13 | .92 | .086 | .68 | .056 | .26 | 5.7 | 1.16 |
|   | 2 | 11 | .94 | .073 | .71 | .051 | .23 | 5.6 | 1.12 |
|   | 3 | 15 | .90 | .099 | .65 | .061 | .29 | 5.8 | 1.19 |
| B | 4 | 7 | .97 | .046 | .80 | .037 | .16 | 5.4 | 1.06 |
|   | 5 | 6 | .98 | .040 | .82 | .033 | .14 | 5.3 | 1.04 |
|   | 6,7 | 5 | .98 | .033 | .85 | .028 | .12 | 5.3 | 1.03 |
|   | 8 | 8 | .96 | .053 | .78 | .041 | .18 | 5.4 | 1.07 |
| C | 9,10,21 | 4 | .99 | .026 | .88 | .023 | .10 | 5.2 | 1.02 |
|   | 11-20,22-26 | 3 | .99 | .020 | .91 | .018 | .07 | 5.2 | 1.01 |
|   | 29 | 2 | .996 | .013 | .94 | .012 | .05 | 5.1 | 1.01 |
|   | 27,28,30-34 | 1 | .996 | .007 | .97 | .006 | .02 | 5.1 | 1.01 |

*Notation*: $N(T)$ - number of target events during $T=404$ years; $\kappa$ - parameter of model (1), $\lambda$ - rate of target events in a zone; $t_- = 2.67$ year; $(n,\tau)$ - errors of the *HK* - optimal strategy, $HK = 1 - n - \tau$, $PG = (1-n)/\tau$; $\sigma/m$ - coefficient of variation of the inter event time (see Appendix).

The prediction errors for $A_c, 0 < c < \infty$ are represented in the $(n,\tau)$ diagram (Fig 2) by a linear segment which connects the error-points related to $A_0$ and $A_\infty$. Because $F_i(s), s > t_+$ is exponential, any prediction in the time zone $s > t_+$ is like a random guessing. Therefore $A_0$ is the most effective alarm among $\{A_c\}$. Hereafter $A_0$ will be treated as the *quasi HK optimal* alarm.

For the Italian model (see Table 2) one has $x_i/\tilde{x}_i = 0.9 - 1.0$; $t_- = 2.67$ years if $\gamma = 0.8$ and $t_+ = 20$ years. In addition, $\lambda_i t_- = 0.01 - 0.1$, i.e., the alarm duration of the quasi *HK* optimal strategy is very short relative to the expected value of the inter-event time, $1/\lambda_i$.

### 4. Numerical results

The statistical estimation of multi-renewal model (1) for Italy (Cinti et al,

2004) is based on the historical/instrumental data for the period T=1600-2003 (Table 1) where one has 136 events of magnitude $\geq 5.5$, and on a seismotectonic zonation of Italy resulting in 34 active zones (Fig.1). To reproduce the basic survival function $1-F_0(s)$, we use the analytical approximation (16) with two parameters $\gamma = 0.8$ and $t_+ = 20\, years$. Using the target event rate $\lambda_i$ in the i-th zone and the basic survival function $1-F_0(s)$ we get the parameter $\kappa_i$ (equation (19)). Finally, $\lambda_i$ comes from Table 1 as the estimate:

$$\hat{\lambda}_i = N_i(T)/T, \qquad (27)$$

where $N_i(T)$ is the number of target events in the i-th zone during the period $T = 404$ years. The rate $\lambda$ defines uniquely the modelled seismic regime in a zone and therefore it is the key parameter in our analysis.

The following indicates the statistical uncertainty of $\{\lambda_i\}$. The 34 zones can be represented by 3 groups (A, B, C) depending on their seismic activity (Fig.1, Table1):

**Table 3.** $N(T)$ data.

| Group | A | B | C |
|---|---|---|---|
| # zones | 3 | 5 | 26 |
| $N_i(T)$ | 11-15 | 5-8 | 1-4 |

i.e., 26 out of 34 zones have only 1 event per 100 or more years. For the most active zones (group A) we can involve small events to compare $N_i(T)$ with their estimates $\hat{N}_i(T)$ based on instrumental data only. Using the Gutenberg-Richter law and the catalogue CSI1.1:1981-2002, $M \geq 2.5$ (http://csi.rm.ingv.it) one has

**Table 4.** G-R estimates of $N_i(T)$

| Zone | 1 | 2 | 3 |
|---|---|---|---|
| $N_i(T)$ | 13 | 11 | 15 |
| $\hat{N}_i(T)$ | 3.5 | 4.5 | 26.9 |

Tables 3-4 show that the problem of correct estimating of $\lambda_i$ exists for all zones.

The uncertainty of $\{\hat{\lambda}_i\}$ is beyond the scope of our further consideration because of two reasons: (i) the operating forecast for Italy does not take into account uncertainty of the multi-renewal model, and (ii) we investigate the predictive ability of the model assuming it to be true.

By Table 3, there is no practically a chance for testing the model in most zones during a reasonable time. Fortunately we have rare opportunity for a theoretical analysis of the prediction.

***Prediction efficiency in the zones***. Table 2 shows the zonal errors $(n_i, \tau_i)$ of the $HK$ optimal strategy and the corresponding efficiency $HK_i = 1 - n_i - \tau_i$. This data can be summarized as follows:

**Table 5.** $HK$ skill score.

| Group of zones | A | B | C |
|---|---|---|---|
| $HK \cdot 100\%$ | 23-29 | 11-18 | 2-10 |
| $n = n_A \cdot 100\%$ | 60-70 | 80-85 | 90-95 |

Approximately $HK_i$ is proportional to $\lambda_i$, namely, $HK_i / \lambda_i t_+ \approx \rho$ where $\rho = 0.4$ for group A, and $\rho = 0.5$ for B and C. On the whole, most zones (31 out of 34) have both low rates $\lambda_i$ and low efficiencies $HK_i$.

To clarify this observation, note that the errors of the $HK$ optimal and the quasi $HK$ optimal strategies are the same within two significant digits. Therefore it is easier to operate with the quasi-optimal strategy. We recall the corresponding alarm rule: in the zone of interest *alarm starts immediately after a target event and is continued during $t_- = 2.67$ years only; in any case the current alarm is stopped by the next event*.

Because $\lambda_i t_-$ are small, the maximum efficiency is reached by successful prediction of secondary events in clusters only. The clustering is not typical for zone of low rate $\lambda_i$; therefore the efficiency has to be low too.

Table 5 contains the following zonal characteristics: $n$ - relative number of failure-to-predict and $n_A$ - relative number of false connected alarms. They are equal and unusually high: $n = n_A = 70 - 90\%$ because the long seismic gaps (>20years) are not infrequent in the model and any prediction of an event during this period is not effective.

Figure 2 shows zonal error diagrams. Each curve contains the linear segment that represent the strategies whose alarm sets are: $A_c = (o, t_-) \bigcup (t_+, t_+ + c)$, $c \geq 0$. The left hand ends of the segments correspond to the quasi $HK$ optimal strategies; they

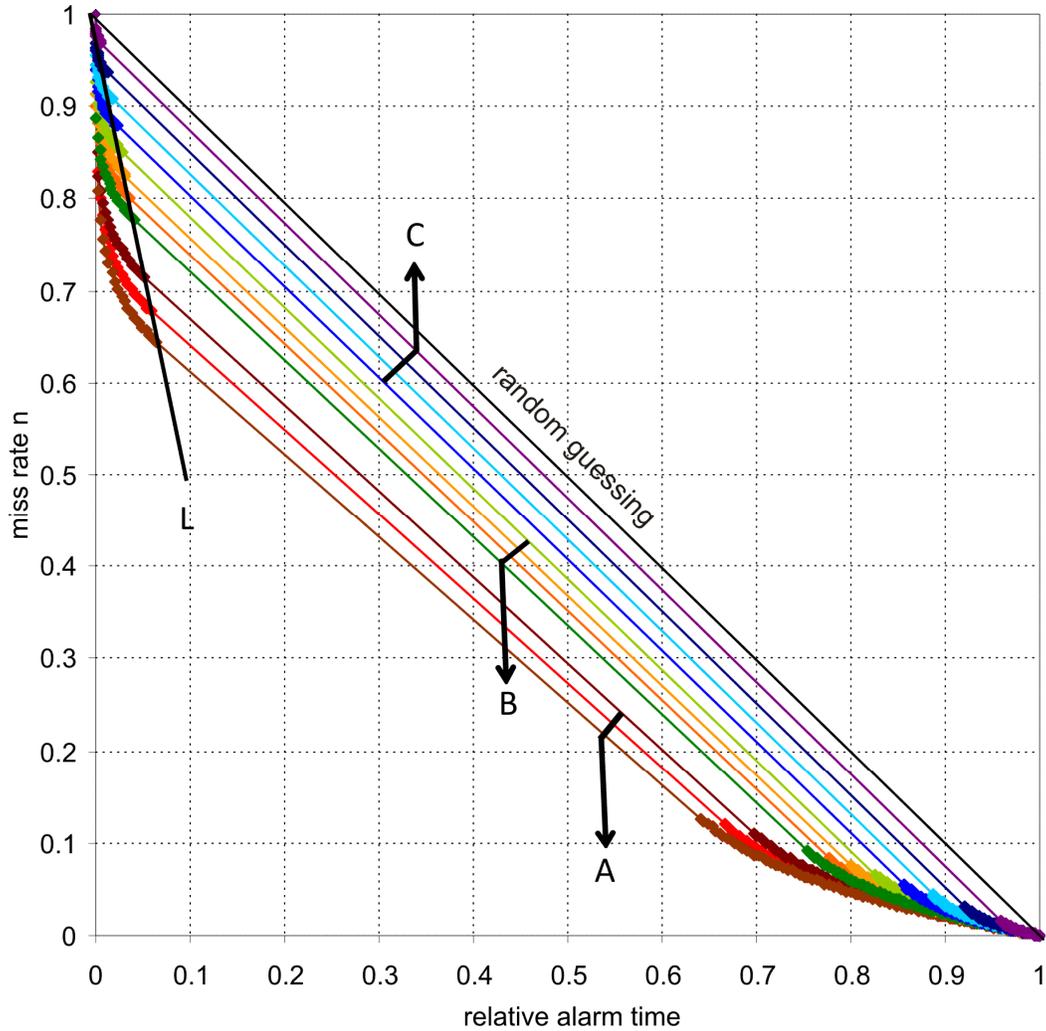

**Figure 2.** Error diagrams for the Italian seismoactive zones.

*Notation.* L - straight line given by the equation $1-n = 5\tau$. A, B, C correspond to the considered groups of the zones. The diagrams of each group are ordered on the plot from the bottom upwards: A {2, 1, 3}, B {8,4,5,(6,7)}, C {(9,10,21), (11-20,22-26), 29, (27,28,30-34)}, where e.g., (6,7) means that the zones 6 and 7 have the same diagram.

line up according to: $(1-n)/\tau \approx (1-\gamma)^{-1}$. (This equality is strict for the limit rate $\lambda_i = 0$). The quantity $(1-n)/\tau$ is known as *probability gain skill score*, *PG*, and is quite often used for comparison of prediction methods (see, e.g., Jordan et al, 2011). In the present case *PG* is high and weakly depends on the zone: 5.7-5.8 (group A) and 5.1-5.2 (group C), (see Table 2). From the other hand, the zonal error diagrams for these groups are quite different. In the limiting case, $\lambda \ll 1$, $PG = 5$ whereas the error diagram is trivial: $n + \tau = 1$. Hence, we have a real example in which $PG$ skill score is non-effective to compare the predictive ability of the model in different zones. The high $PG$ values mean only that a short alarm after any large event can be useful to predict next large aftershock in the same zone. But this utility highly depends on the rate of main events.

To avoid confusion, note that the probability gain term is used in two ways: (i) as a time dependent quantity to characterize the hazard locally and (ii) as a 'cumulative' quantity (skill score) to characterize a prediction strategy as a whole. The local $PG$ is the effective base to optimize the predictions (Molchan ,1997).

***Total prediction efficiency.*** Table 6 shows the prediction characteristics ($HK_\Sigma$ and $n_\Sigma$) of the $HK$-optimal strategy for the different groups of zones A, A+B and A+B+C.

**Table 6.** Total $HK$ skill score.

| Group of zones | A | A+B | A+B+C |
|---|---|---|---|
| $HK_\Sigma \cdot 100\%$ | 26 | 21 | 15 |
| $n_\Sigma \cdot 100\%$ | 68 | 74 | 82 |
| $\hat{n}_\Sigma \cdot 100\%$ | 77 | 80 | 86 |

The theoretical prediction of the miss rate $n_\Sigma$ is more optimistic than the empirical one, $\hat{n}_\Sigma$. The estimate $\hat{n}_\Sigma$ is based on the data for the period $T = 404$ years and corresponds to the alarm window $(0, t_-)$. Probably, the divergence between $n_\Sigma$ and $\hat{n}_\Sigma$ is a result of aftershock dependence because the estimates $\hat{n}_\Sigma$ are equal for any $t_- = 0.5\text{-}4$ years. This effect is shown up after the model fitting and is more evident in group A where we have 50% of the catalogue's aftershocks.

Formally, $HK_\Sigma$ is the weighted-mean of the zonal $HK$-values (see (25)). Therefore, the total score $HK_\Sigma$ can distort the predictive ability of the model in the zones with extreme $HK$-values. For example, the typical $HK$ - values are 23-29% for the zones of group A and 2-7% for group C whereas $HK_\Sigma = 15\%$ for all zones together. In other words, *the zones of low activity act as a noise when we characterise the prediction ability of the model as a whole.*

Figure 3 shows the main prediction characteristics ($n_\Sigma(p), \tau_\Sigma(p), HK_\Sigma(p)$) for the forecasting strategy { $h_i(s) > p$ } in A, A+B, and A+B+C. Instead of $p$ we use

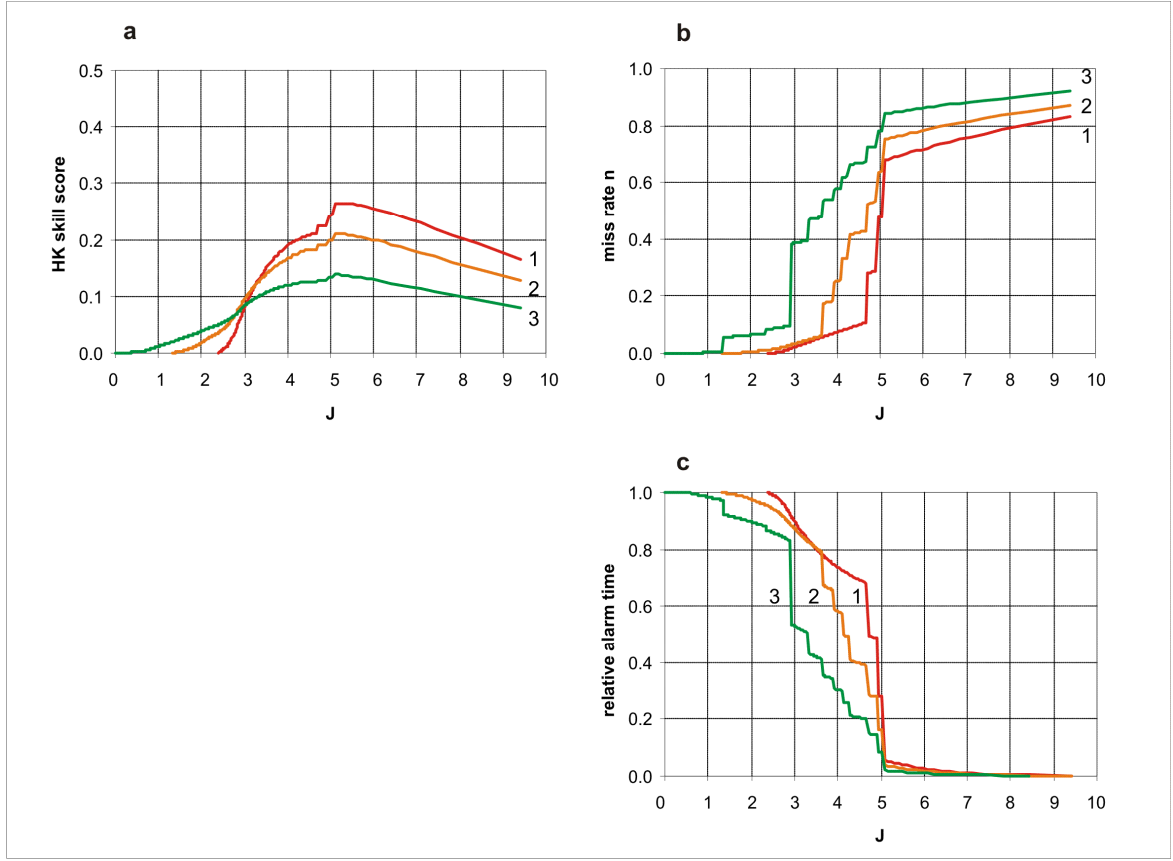

**Figure 3.** Prediction characteristics: a) $HK_\Sigma(p)$, b) $n_\Sigma(p)$, c) $\tau_\Sigma(p)$ vs. $J = \log_2 p + 10$ (hazard level) for forecasting strategy (11) in three groups of zones. Notation: 1 - group A, 2 - group A+B, and 3 - group A+B+C.

the following log- scale: $J = \log_2 p + 10$.

The zonal error diagrams for elements of the group A are similar. Therefore the total prediction characteristics of the group A represent the individual zones as well. In cases of A+B and A+B+C the characteristics show again the noise influence of low rate zones.

## 5. Discussion

**Fitting of the multi-renewal model.** We consider here the key elements of the statistical estimation of the seismicity model (1,16).

*Regionalization* (decomposition of the region into homogenous tectonic zones). Generally this procedure is not uniquely defined. Therefore it is poorly compatible with independence of target events in the zones. To illustrate this suppose that we wish to refine the original regionalization by subdividing a zone $G_0$ into two parts, $G_{01}$ and $G_{02}$ (compare e.g. Cinti et al (2004) and Faenza & Marzocchi (2010)). The original zoning assumes the events to occur according to a renewal model. The new zoning will again assume as a minimum that the events in $G_{01}$, $G_{02}$ are independent. Then, given some minimal regularity conditions, the recurrence of target events in $G_{01}$ can also be a renewal process if and only if the events in $G_{01}$ are poissonian (Daley& Vere-Jones, 1988, p.79). In short, the assumption of the renewal property for the occurrence of events in an area and in its sub-parts is not self-contradictory for the Poisson model only.

Combining two zones with independent (non-poissonian) renewal processes, the originally independent inter-event times of one zone are spitted by events from other zone into dependent parts, thereby violating the condition for their independence. The hazard function $h(s)$ will then be overestimated for small $s$ due to larger numbers of small intervals. On the contrary, a splitting of the process into two sequences leads to an opposite effect, a smaller hazard function for small $s$.

*The independence of inter-event times*. The property of independence between events in different zones do not influence the efficiency of prediction $HK$. However, the assumption of independence for the population of the past inter-event times is the key element for estimating the basic hazard function and therefore it affects the forecasting probabilities as a whole.

*The proportionality of hazard functions* (PHF). From the statistical point of view, the PHF assumption can distort the distribution of inter-event intervals, because the basic hazard function for small times is generated by a small number of active zones, whereas for moderately large times it is generated by a large number of zones with low seismicity and poor clustering.

The PHF implies the equality of hazard functions in zones with identical rates of target events independently of the tectonics involved. This fact has the following consequence.

**Regression in the Proportional Hazard (PH) model.** The PH model allows reducing the number of unknown parameters by using the following regression

$$\ln \kappa_i = \sum_{\alpha=0}^{m} z_\alpha^{(i)} \beta_\alpha \ . \tag{28}$$

Here $(\beta_0,...\beta_m)$ are new unknown parameters, and $(z_0^{(i)},...z_m^{(i)})$ are some tectonic/physical characteristics of zone $G_i$. For example, Cinti et al (2004) use the following z-components: $z_0$ - log-rate of the main target events (the term 'main' is related to the time-space aftershock window: 3months×30km×30km), $z_1$ - prevalent stress regime, $z_2$ - homogeneity of the stress orientation, $z_3$ - fault code, $z_4$ - topographic homogeneity.

Regression (28) makes sense in case of large number of zones, $N_\kappa \gg m$. However, scarcity of data is a serious obstacle for application of regression (28) because the number of significantly different estimates of $\lambda$ in the zones, $N_\lambda$, is small.

Indeed, by (19), $\ln \kappa$ is a function of $\lambda$. Therefore the requirement $m \ll N_\lambda$ is more reasonable than $m \ll N_\kappa$. The estimate of $N_\lambda$ come from Table 2: $N_\lambda < 10$. The zones without aftershocks during period $T$ have equal point estimates of $\lambda$ and $z_0$. Therefore for such zones, one has

$$\sum_{\alpha=1}^{m}(z_\alpha^{(i)} - z_\alpha^{(j)})\beta_\alpha = 0, \qquad \text{given } \lambda_i = \lambda_j. \tag{29}$$

If the matrix of linear system (29) has rank $m$, then, obviously, $(\beta_1,...,\beta_m) = 0$.

The zones with equal number of target events $N(T)$ and no aftershocks among them form the following groups {11-20, 22-24} and {27-28, 30-34}. Therefore, system (29) contains at least 18 independent equations relative to $m = 4$ unknown parameters. It is enough to conclude that regression (28) is degenerate (the $z$-data from Cinti et al (2004) supports this conclusion).

Note that Cinti et al, (2004) empirically came to the one-factor regression model (28). The unique factor is $z_0$: it has to dominate in (28) because $\kappa_i$ and $z_0^{(i)}$ are interpreted similarly (see section 3). Hence, the regression possibilities of the Cox model in the considered case are useless.

**Forecasting or prediction?** There are too many aspects of the question to be uniquely answered. Formally, the probability $p_i(t|\Delta)$ of a target event in a space-time volume $G_i \times \Delta$ can be transformed into prediction (0-1 expression) by fixing a threshold $p_i^0$. In reality, we have to restrict oneself in case of large events prediction by few thresholds because of scantiness of learning data. To guarantee a reliability of prediction statements, the choice of the thresholds has to be realized at the research stage but not at the operational one (see for comparison Jordan et al., 2011). Hence, a prediction algorithm for large events can be considered as a forced restriction of the corresponding forecasting method.

The choice of the thresholds { $p_i^0$ } depends on the prediction goal for the whole region (see e.g. (8) where the weight of the alarm time unite, '$a$', may depend on $G_i$). A forecasting, restricted by { $p_i(t|\Delta)$ } information only, automatically introduces a common probability scale ( $p_i^0 = p^0$ ) to compare the hazards in the zones. On the contrary, a prediction is usually based on a relative probability scale (see section 2).

The information { $p_i(t|\Delta)$ }, which can't be reliable enough, is important during the testing stage of a forecasting method and must be completed by a goal function at the application stage. We remind the history of taking decision in favor of the Parkfield experiment in 1988 to predict a characteristic event on the San Andreas fault. Using the multi-renewal model with the log-normal distribution, the Working Group on California Earthquake Probabilities (1988) found for 8 fault segments the probability of a characteristic event during the period $\Delta$ = (1988-2018), namely, 90% in Parkfield vs. <40% for the other segments. According to this approach, the Parkfield segment turned out to be permanently the most hazardous. On the other hand, according to the *HK*-optimal prediction, five of the eight fault segments should be in a state of permanent alert after 1988. The collision was owing to the choice of a common threshold and the long range prediction horizon $\Delta$=30 years. The relative value of the horizon, $\lambda_i \Delta$, is large for Parkfield (1.5) and very small for the other faults (0.1-0.2) . In such case the probability of target event is highest for Parkfield *a priory* (see more in Molchan, 2003). A similar effect we can see on the probability map by Cinti et al (2004) for $M \geq 5.5$ in Italy for the next 10 years.

**6.Conclusion**

Accepting a multi-renewal model for prediction purposes we have to analyze the predictive ability of the simplest information: the elapsed time since the last target event in zones of interest. The model for Italy includes two reasonable properties of target events: clustering at small time scales and the memoryless property at large time scales (>20 years). As a result, the clustering is the main information basis in the prediction of $M \geq 5.5$ events. This basis is not effective for most zones, 26 out of 34, where only one event occurred in 100 or more years.

To optimize the $HK$ skill score (the relative rate of events predicted not accidentally) it is enough to use the following very simple alarm rule: in any of 34 zones *the next target earthquake is expected to occur within $t_- = 2.67$ years starting from the last event in the zone*. This rule excludes any additional calculations and ensures (provided that the model is true) the following HK values: ~26% for 3 most active zones (group A) and 2-10 % for 26 least active zones (group C).

For the C-zones, where the rate of target events is low: 0.0025-0.01, the $HK = 10\%$ means that only one nonrandom success is possible per 1000-4000 years. In addition to $HK$, one can see that the number of false connected alarms per one event in a zone is unusually high: ~0.7 (group A) and 0.8-0.95 (group C).

In the considered model the probability gain score is high for all zones ($PG$ =5-6) and therefore unproductive for comparison of predictions.

On the whole, our analysis supports the following conclusion: *4 out of 34 zones (1-3,8) are of some interest for testing the model in prediction of $M \geq 5.5$*. Any changes in regionalization imply a new statistical estimation of the basic hazard function. In framework of optimization of the $HK$ skill score the complicated statistical procedure proposed by (Cinti et al., 2004) can be simplified by fitting of the alarm window parameter $t_-$ only.

The standardized approach to testing of forecasting methods (Schorlemmer et al, 2007) does not work in prediction of large earthquakes (Marzocchi & Zechar, 2011, Molchan, 2012). An alternative way is a specific approach to the problem depending on the method and the data. The following are such examples: the theoretical analysis of the Italian predictions presented here, and the statistical analysis of M8 algorithm with predictions of M:8-8.5 events worldwide (Molchan & Romashkova, 2010).

**Acknowledgments**
We are thankful to A. Peresan for helpful discussion of the forecasting problems in Italy and very useful comments; to F. Cinti for kind providing us with the seismotectonic regionalization of Italy. One of the authors (G.M.) is grateful to G.Panza for the invitation to University of Trieste for a work on the paper.**References**

Cinti ,F.R., Faenza, L., Marzocchi,W., P. Montone,P., 2004. Probability map of the next M > 5.5 earthquakes in Italy. *Geochemistry, Geophysics, Geosystems* , electronic j. of the earth sci. , AGU and the Geochemical Soc., 5( 11), 1-15, doi:10.1029/2004GC00072.

Daley, D.J. and Vere-Jones,D., 1988. *An Introduction to the Theory of Point Processes*, Springer, New York.

Faenza, L., Marzocchi,W., Boschi,E., 2003. A nonparametric hazard model to characterize the spatio-temporal occurrence of large earthquakes: An application to the Italian catalogue, *Geophys. J. Int*., 155(2), 521–531.

Faenza, L., Marzocchi**,**W., 2010. The Proportional Hazard Model as applied to the CSEP forcasting area in Italy. *Annales of Geophysics*, 53(3), 77-84, doi: 10.4401/ag-4759.

Jordan, T, Chen,Y., Gasparini,P., Madariaga,R., Main,I., Marzocchi,W., Papadopoulos,G., Sobolev,G., Yamaoka,K., Zschau,J., 2011. ICEF Report. Operational earthquake forecasting: state of knowledge and guidelines for utilization. *Annals Geophys,* 54(4), doi:10.4401/ag-5350.

Kalbfleisch, J., Prentice,R., 2003. *The statistical analysis of failure time data*. J.Wiley&Sons Inc., N.Y., 2nd edn.

Marzocchi, W. and Zechar, J.D. Earthquake forecasting and earthquake prediction: different approaches for obtaining the best model. *Seism. Res. Lett.* 82(3), 442-448. doi:10.1785/gssrl.82.3.442

Molchan G., 1990. Strategies in strong earthquake prediction. *Phys. Earth and Planet. Inter*., 61( 1-2), 84-98.

Molchan G., 1991. Structure of optimal strategies in earthquake prediction. *Tectonophysics,* 193, 267-276.

Molchan, G., 1997. Earthquake Prediction as a Decision-making Problem. *Pure Appl. Geophys*., 149, 233-247.20

Molchan, G., 2003. Earthquake prediction strategies: a theoretical analysis. In *Nonlinear dynamics of the lithosphere and earthquake prediction*, pp.209-237, eds. Keilis-Borok,V.I. and Soloviev, A.A. , Springer.

Molchan, G., 2010. Space-Time Earthquake Prediction: the Error Diagrams. *Pure Appl. Geophys.*, 167(8-9), 907-917, doi: 10.1007/s00024-010-0087-z

Molchan, G., 2012. On the Testing of Seismicity Models. *Acta Geophysica*, 60(3), 624-637, doi: 10.2478/s11600-011-0042-0

Molchan,G. and Keilis-Borok, V.I., 2008. Earthquake Prediction: probabilistic aspect. *Geophys. J. International*, 173(3), 1012-1017.

Molchan G., Romashkova,L., 2010. Earthquake Prediction analysis based on empirical seismic rate: the M8 algorithm. *Geophys. J. Int.*, 183, 1525-1537.

Schorlemmer,D., M.Gerstenberger, S.Wiemer, D.Jeckson, and D.Rhoades, 2007, Earthquake likelihood model testing, *Seismol. Res. Lett.* 78, 17-29, doi: 10.1785/gssrl.78.1.17.

Working Group on California Earthquake Probabilities ,1988. Probabilities of large earthquakes occurring in California on the San Andreas fault. *U.S. Geol. Survey Open File Rep.*, 88-398.

**Appendix**

*The coefficient of variation $I = \sigma/m$.*

Consider the distribution $F(s) = 1 - \exp\{-\kappa \int_0^s h_0(u)du\}$, where $h_0(s)$ is given by (12). Denoting by $m_1, m_2$ the first two moments of *F*, we get

$$m_2/m_1^2 = 2[\Gamma(x|2\bar{\gamma})x^{-2\bar{\gamma}}\bar{\gamma} + (1+x)e^{-x}](\tilde{x}/x)^2, \qquad (A1)$$

where $x = \kappa t_+$, $\tilde{x} = \lambda t_+$, $\bar{\gamma} = (1-\gamma)^{-1}$ and $\Gamma(x|\alpha)$ is the incomplete Gamma function. Hence, $I = (m_2/m_1^2 - 1)^{1/2}$.

According to Table 2, *I*=1-1.2 for the Italian model. Usually the relation *I* ~1 is considered as an indicator of a poor forecasting ability of the renewal model.